\documentclass[twocolumn]{aastex63}

\received{XXX}
\revised{XXX}
\accepted{XXX}
\submitjournal{AJ}

\shorttitle{Metal-poor stars in $\omega$ Cen}
\shortauthors{Johnson et al.}

\begin{document}

\title{The Most Metal-poor Stars in Omega Centauri (NGC 5139)\footnote{This 
paper includes data gathered with the 6.5 meter Magellan Telescopes located at 
Las Campanas Observatory, Chile.}}

\correspondingauthor{Christian I. Johnson}
\email{chjohnson1@stsci.edu}

\author[0000-0002-8878-3315]{Christian I. Johnson}
\affiliation{Space Telescope Science Institute, 3700 San Martin Drive, 
Baltimore, MD 21218, USA}

\author{Andrea K. Dupree}
\affiliation{Center for Astrophysics $\vert$ Harvard \& Smithsonian, 60 Garden Street, MS-15, Cambridge, MA 02138, USA}

\author{Mario Mateo}
\affiliation{Department of Astronomy, University of Michigan, Ann Arbor, MI 48109, USA}

\author{John I. Bailey, III}
\affiliation{Department of Physics, UCSB, Santa Barbara, CA 93016, USA}

\author{Edward W. Olszewski}
\affiliation{Steward Observatory, The University of Arizona, 933 N. Cherry Avenue, Tucson, AZ 85721, USA}

\author{Matthew G. Walker}
\affiliation{McWilliams Center for Cosmology, Department of Physics, Carnegie Mellon University, 5000 Forbes Avenue, Pittsburgh, PA 15213, USA}



\begin{abstract}

The most massive and complex globular clusters in the Galaxy are thought to
have originated as the nuclear cores of now tidally disrupted dwarf galaxies,
but the connection between globular clusters and dwarf galaxies is tenuous
with the M54/Sagittarius system representing the only unambiguous link.  The 
globular cluster Omega Centauri ($\omega$ Cen) is more massive and chemically 
diverse than M 54, and is thought to have been the nuclear star cluster of 
either the Sequoia or Gaia-Enceladus galaxy.  Local Group dwarf galaxies 
with masses equivalent to these systems often host significant populations of 
very metal-poor stars ([Fe/H] $<$ $-$2.5), and one might expect to find such 
objects in $\omega$ Cen.  Using high resolution spectra from Magellan-M2FS, we 
detected 11 stars in a targeted sample of 395 that have [Fe/H] ranging from 
$-$2.30 to $-$2.52.  These are the most metal-poor stars discovered in the 
cluster, and are 5$\times$ more metal-poor than $\omega$ Cen's dominant 
population.  However, these stars are not so metal-poor as to be unambiguously
linked to a dwarf galaxy origin.  The cluster's metal-poor tail appears to 
contain two populations near [Fe/H] $\sim$ $-$2.1 and $-$2.4, which are very 
centrally concentrated but do not exhibit any peculiar kinematic signatures.  
Several possible origins for these stars are discussed.

\end{abstract}

\keywords{globular clusters: general, globular clusters: individual: Omega Centauri, stars: abundances, stars: Population II, Astrophysics - Solar and Stellar Astrophysics}


\section{Introduction}

Omega Centauri ($\omega$ Cen) possesses an elegant complexity that is unmatched
by any other Galactic globular cluster.  While most globular clusters exhibit
metallicity dispersions of $\sim$ 0.05 dex or less \citep[e.g.,][]{Carretta09,
Bailin19}, stars in $\omega$ Cen span at least a factor of 100 in 
[Fe/H]\footnote{[A/B] $\equiv$ log(N$_{\rm A}$/N$_{\rm B}$)$_{\rm star}$ $-$
log(N$_{\rm A}$/N$_{\rm B}$)$_{\sun}$ for elements A and B.} and are 
distributed into at least 5 distinct populations with unique metallicities 
\citep[e.g.,][]{Suntzeff96,Norris95,Norris96,Villanova07,Johnson10,Marino11,
Pancino11}.  Each distinct metallicity group can be further partitioned into at
least 2-3 sub-populations with variable light element chemistry, and $\omega$ 
Cen may host as many as 15 unique stellar populations \citep{Bellini17}.  

The origin and enrichment history of $\omega$ Cen is currently an open
question, but several lines of evidence suggest that the cluster may be the
remnant core of a now disrupted dwarf galaxy.  For example, $\omega$ Cen is
the brightest and most massive globular cluster in the Milky Way, but has a
strong retrograde orbit \citep{Dinescu99} that may be associated with the
capture and disruption of the ``Sequoia" galaxy discussed in \citet{Myeong19}.
Alternatively, $\omega$ Cen could have once been the nuclear star cluster of
the more massive Gaia-Enceladus galaxy \citep{Massari19}.  \citet{Bekki03} used
dynamical models to show that a compact nuclear core like $\omega$ Cen could 
survive such a disruption event, and in fact the long tidal stream recently 
found by \citet{Ibata19}, which stretches several degrees along the cluster's 
orbit, seems to support an accretion origin \citep[see also][]{Simpson19}.  We 
also note that $\omega$ Cen shares many physical and chemical characteristics 
with M 54 \citep[e.g.,][]{Carretta10}, which is the most massive cluster in
the Sagittarius dwarf galaxy that is currently being tidally destroyed by the 
Milky Way.

In terms of size and luminosity, $\omega$ Cen and M 54 lie at the intersection
between globular clusters, nuclear star clusters, and Ultra Compact Dwarfs
\citep[e.g.,][]{Mackey05,Georgiev09,Tolstoy09}, and may be prototypes for
``iron-complex" clusters\footnote{Note that iron-complex clusters are generally
the same as the ``anomalous" and ``Type II" clusters identified by
\citet{Marino15} and \citet{Milone17}, respectively.} that host multiple
generations of stars with different light and heavy element abundances
\citep[e.g.,][]{Yong14,Johnson15,Marino15,DaCosta16,Milone17}.  Although a 
definitive connection between iron-complex clusters and dwarf galaxy nuclei has
not yet been established \citep[e.g., see][]{DaCosta16}, one possible 
investigation path is to search for cluster members that may have originally
been part of the progenitor galaxy's field star population.  Such stars could
be identified as having chemistry that is inconsistent with known globular
cluster patterns.

For $\omega$ Cen, the two most likely populations to have originated as 
dwarf galaxy field stars are those at the highest ([Fe/H] $\ga$ $-$1) and 
lowest ([Fe/H] $\la$ $-$2) metallicities.  On the high metallicity end, stars
exhibit relatively small light element abundance variations and critically
show an O-Na correlation rather than the expected anti-correlation
\citep{Johnson10,Marino11,Pancino11_mr}.  However, several authors have noted 
that this particular pattern may be attributed to self-enrichment driven by
an unusually long star formation history \citep{Johnson10,DAntona11,Marino11}.
The iron-complex clusters M 2 \citep{Yong14} and NGC 6273 \citep{Johnson15,
Johnson17} also contain minority populations of metal-rich stars with small
star-to-star abundance variations, but in these cases no clear O-Na 
correlations are present.  For all three clusters the origins of their 
metal-rich populations remain ambiguous, and it is not yet possible to 
differentiate whether these stars were captured from a progenitor galaxy or
simply trace prolonged chemical enrichment.

The most metal-poor $\omega$ Cen stars have a greater potential to 
unambiguously link the cluster with a dwarf galaxy formation environment.
For example, Milky Way and extragalactic globular cluster systems exhibit a 
clear metallicity floor of [Fe/H] $\sim$ $-$2.3 to $-$2.5 
\citep[e.g.,][]{Beasley19}, which suggests that any $\omega$ Cen stars more
metal-poor than this limit likely did not originate in the cluster.  The 
metallicity distribution function of the Sequoia galaxy, proposed by 
\citet{Myeong19} as the progenitor system for $\omega$ Cen, is not currently 
well-constrained, but several estimates indicate that it likely had a mass in 
excess of 10$^{\rm 7}$ M$_{\rm \odot}$ \citep[e.g.,][]{Bekki19,Myeong19}.  The 
dwarf galaxy mass-metallicity relation from \citet{Kirby13} suggests that such 
a system should have a mean [Fe/H] $\sim$ $-$1.5, but if $\omega$ Cen formed
in a 10$^{\rm 10}$ M$_{\rm \odot}$ system, such as Gaia-Enceladus 
\citep{Helmi18}, then the host galaxy's mean metallicity may have been closer
to [Fe/H] $\sim$ $-$0.5.  In either case, the metallicity distribution 
functions of most existing dwarf galaxies exhibit long metal-poor tails that
reach at least 1-2 dex lower than their mean [Fe/H] values 
\citep[e.g.,][]{Starkenburg10,Kirby11,Leaman13}, which suggests that any dwarf
galaxy massive enough to host $\omega$ Cen likely had $\sim$ 1-10$\%$ of its
stars with [Fe/H] $\la$ $-$2.5.

Previous analyses of $\omega$ Cen's red giant branch (RGB) and subgiant branch
(SGB) populations have so far failed to find any stars with [Fe/H] $\la$
$-$2.25, which is still within the metallicity realm of Galactic globular
clusters.  Recent RR Lyrae investigations by \citet{Bono19} and 
\citet{Magurno19} found a small number of stars with [Fe/H] $\la$ $-$2.5; 
however, the mean cluster metallicities of these studies are $\sim$ 0.2-0.3 dex 
lower than spectroscopic analyses of RGB stars \citep[e.g.,][]{Johnson10}, 
which suggests their true metallicity floor may be closer to [Fe/H] $\sim$ 
$-$2.3.  

Broad-band color-magnitude diagram analyses alone are unlikely to find very 
low metallicity stars in $\omega$ Cen because the RGB evolutionary sequences of
old populations become difficult to distinguish in color below [Fe/H] $\sim$ 
$-$2 \citep[e.g., see Figure 7 of][]{Simpson18}.  Furthermore, the isochrones 
of very metal-poor RGB stars in $\omega$ Cen may overlap in color and magnitude
space with asymptotic giant branch (AGB) sequences as well.  Many of the bluer 
$\omega$ Cen stars, especially in regions where RGB and AGB stars of different 
metallicity can mix, do not yet have spectroscopic [Fe/H] determinations, and 
as a result the most metal-poor stars in the cluster may have been missed by 
previous surveys.  Therefore, we derive spectroscopic [Fe/H] measurements for 
395 of the bluest RGB stars in $\omega$ Cen to search for the cluster's most 
metal-poor constituents that may link the system to formation in a dwarf galaxy
environment.

\section{Observations and Data Reduction}

\begin{figure*}
\includegraphics[width=\textwidth]{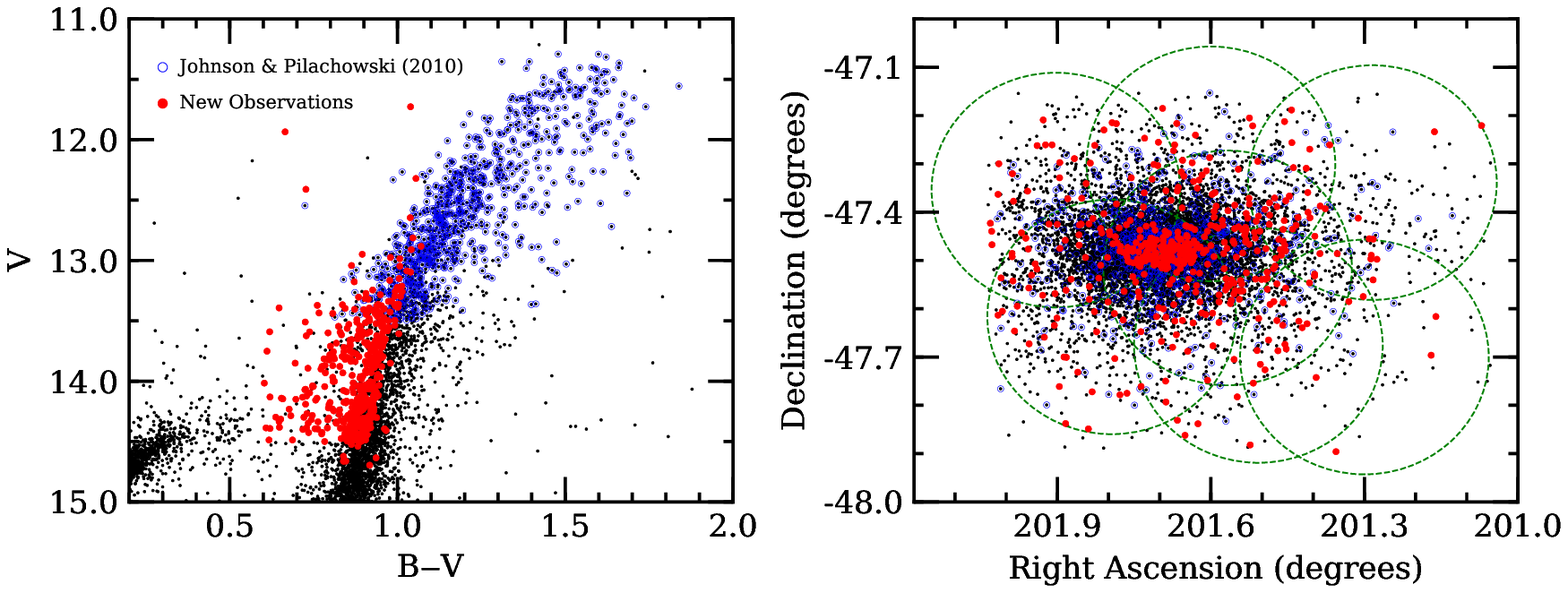}
\caption{The left panel shows a V versus B$-$V color-magnitude diagram for all
stars with membership probabilities $>$ 70$\%$ in the \citet{vanLeeuwen00}
catalog.  Stars observed in \citet{Johnson10} and the present study are
indicated with open blue circles and filled red circles, respectively.  The
right panel uses the same symbols to illustrate the sky coordinates of the
targets observed in both spectroscopic studies.  The M2FS configuration fields
are identified with green dashed circles.}
\label{fig:cmd_plot}
\end{figure*}

All spectroscopic data for this project were obtained using the 
Michigan-Magellan Fiber System \citep[M2FS;][]{Mateo12} and MSpec spectrograph 
mounted on the Magellan-Clay 6.5m telescope at Las Campanas Observatory.  The
spectra were acquired in three runs via the M2FS queue that spanned 2015 
February 19, 21, 22, 23, and 25 for the first set, 2015 March 01, 02, and 04 
for the second set, and 2015 July 18 and 20 for the third set.  Ten different 
fiber configurations were used for seven unique pointings, which are 
illustrated in the right panel of Figure~\ref{fig:cmd_plot}.  Three of the 
fields that overlapped with the cluster core included two separate fiber 
configurations in order to maximize the number of observed stars.  The 
remaining 4 fields each only included a single fiber configuration.

Target stars were selected using the photometry and coordinates from 
\citet{vanLeeuwen00}.  Only objects identified by \citet{vanLeeuwen00} as
having membership probabilities $>$ 70$\%$ were targeted, and we prioritized
stars on the blue half of the RGB with V magnitudes between about 13.5 and
14.5 (see Figure~\ref{fig:cmd_plot}).  Although most stars in our sample are 
first ascent RGB stars, we also included some AGB and red horizontal branch 
(HB) stars since the isochrones for very metal-poor RGB stars are difficult
to distinguish from these later evolutionary stages in more metal-rich 
populations.  Including all configurations, fibers were placed on 458 unique 
targets spanning a variety of cluster radii (see Figure~\ref{fig:cmd_plot};
Table 1); however, we were only able to measure [Fe/H] values for 395 targets. 
The remaining 63 targets had poor S/N spectra, were heavily blended 
spectroscopic binaries, or were cases in which we could not converge to a 
stable model atmosphere solution.

\begin{figure*}[ht!]
\includegraphics[width=\textwidth]{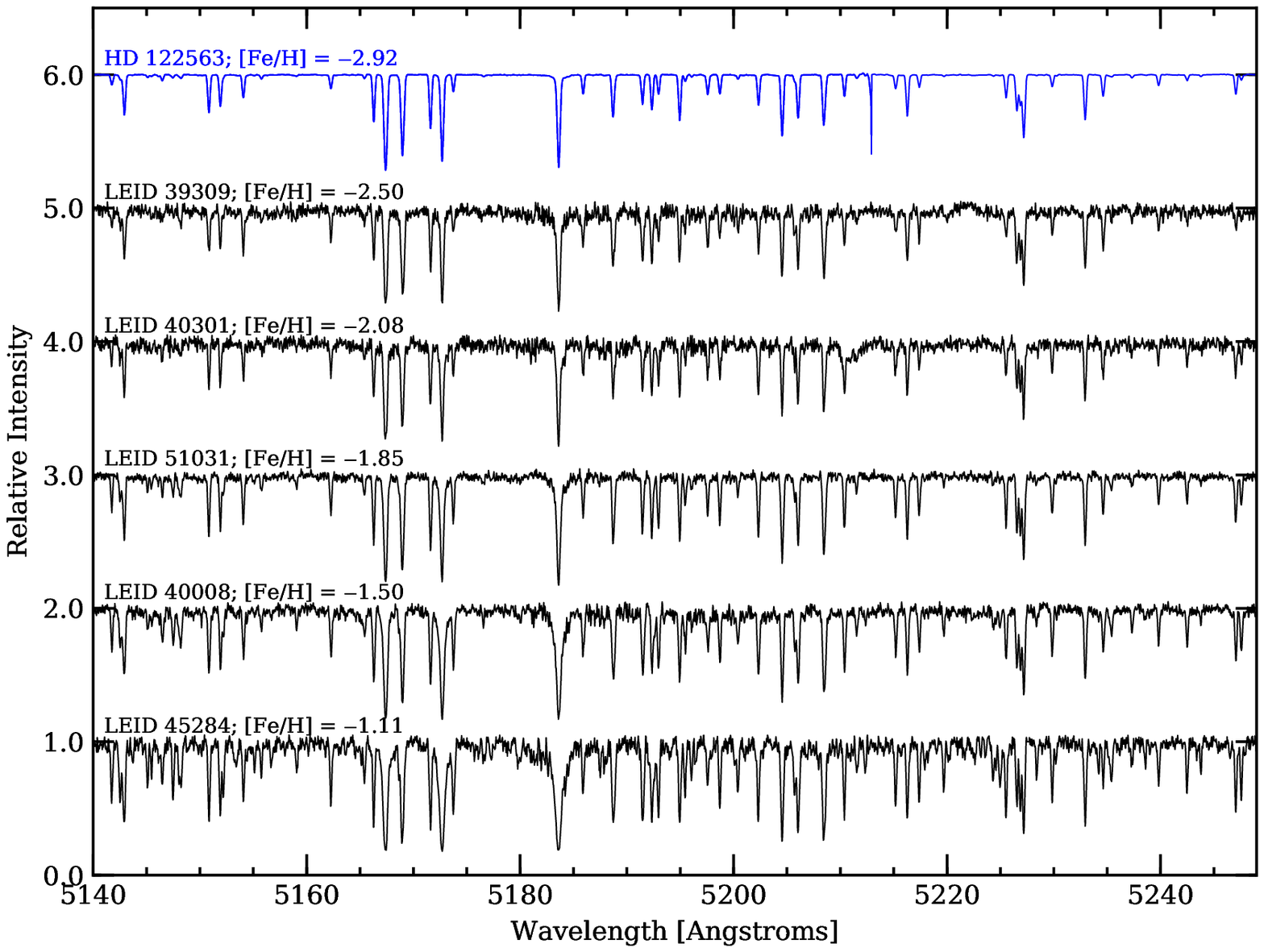}
\caption{Sample M2FS spectra (black lines) are shown for $\omega$ Cen stars 
that span [Fe/H] = $-$2.50 to $-$1.11 in increments of approximately 0.35 dex.
A smoothed and resampled MIKE spectrum of the metal-poor giant HD 122563 is 
shown in blue for comparison.  Note that all of the stars illustrated here have
similar temperature and surface gravity values.  A simple line strength 
comparison shows that LEID 40301 and LEID 39309 are clearly more metal-poor than
LEID 51031, which represents the dominant $\omega$ Cen population, but do not
reach the very low level of [Fe/H] = $-$2.92 found by \citet{Afsar16} for 
HD 122563.}
\label{fig:spec_comp2}
\end{figure*}

All observations utilized the ``Mg$\_$Wide" filters that permit continuous
wavelength coverage from approximately 5125-5410 \AA.  This particular setup 
was chosen because we are searching for stars with [Fe/H] $\la$ $-$2.3 and the 
Fe lines in this region remain detectable in the spectra of T$_{\rm eff}$ 
$\sim$ 5000 K RGB stars down to at least [Fe/H] = $-$4 \citep[e.g., see Figure 
1 of][]{Frebel15}.  The ``red" and ``blue" spectrographs employed identical 
configurations with 1 $\times$ 2 (dispersion $\times$ spatial) CCD binning, 
95$\mu$m slits, a four amplifier slow readout, and provided a 
resolving power of R $\equiv$ $\lambda$/$\Delta$$\lambda$ $\approx$ 32,000.  
Since the Mg$\_$Wide filter spans four consecutive orders, we were only able
to observe a maximum of 24 targets per spectrograph at a time.  Each fiber
configuration was observed for one hour with a set of 3$\times$1200 second
exposures.

The data reduction procedure generally followed the methods outlined in 
\citet{Johnson15_AGB}.  Briefly, the IRAF\footnote{IRAF is distributed by 
the National Optical Astronomy Observatory, which is operated by the 
Association of Universities for Research in Astronomy, Inc., under cooperative 
agreement with the National Science Foundation.} \emph{CCDPROC} routine was
used to apply the bias corrections, trim the overscan regions, and subtract
dark current for each amplifier image.  The \emph{IMTRANSPOSE} and 
\emph{IMJOIN} routines were then used to align and combine the four amplifier
images of every exposure to create sets of monolithic 2K$\times$4K images.  The
\emph{DOHYDRA} task was utilized to handle aperture identification, aperture
tracing, flat-fielding, wavelength calibration via ThAr comparison spectra, 
scattered light and cosmic ray removal, throughput corrections, and spectrum
extraction.  Median sky spectra were generated for each order on each night 
from the designated sky fiber data, and were subtracted from the appropriate
science spectra.  The sky subtracted data were then corrected for heliocentric
velocity variations, continuum normalized, and median combined.  Sample 
combined spectra for several stars with different [Fe/H] but similar temperature
and surface gravity values are provided in Figure~\ref{fig:spec_comp2}.  
Signal-to-noise (S/N) ratios for the combined spectra ranged from $\sim$ 20-50 
per reduced pixel.

\section{Data Analysis}

\subsection{Model Atmospheres}

For consistency with the \citet{Johnson10} temperature/gravity scale,
model atmosphere parameters were determined for each star using dereddened B, 
V, and K$_{\rm S}$-band photometry from \citet{vanLeeuwen00} and the Two Micron
All Sky Survey \citep[2MASS;][]{Skrutskie06}.  Although several authors note 
the existence of minor ($\Delta$ $\sim$ 0.02 mag.) differential reddening near
the cluster core \citep[e.g.,][]{Calamida05,vanLoon07,McDonald09}, we assumed a
constant E(B$-$V) = 0.11 \citep{Calamida05} across the cluster and also 
utilized the relation E(V$-$K$_{\rm S}$)/E(B$-$V) = 2.70 from \citet{McCall04}.
For target stars with clear 2MASS matches, effective temperatures 
(T$_{\rm eff}$) were determined using the V$-$K$_{\rm TCS}$ color-temperature 
relation from \citet{Alonso99}\footnote{Note that the 2MASS K$_{\rm S}$-band 
photometry were transformed onto the K$_{\rm TCS}$ system using the 
transformations listed in \citet{Johnson05}.}.  In the few cases where clear 
2MASS matches could not be obtained we used the B$-$V color-temperature 
relation from \citet{Alonso99} to derive T$_{\rm eff}$ values.  The estimated
T$_{\rm eff}$ uncertainty is approximately 30 K for temperatures derived from
V$-$K$_{\rm TCS}$ and 130 K for B$-$V, based on the scatter of the empirical
color-temperature relations in \citet[][see their Table 2]{Alonso99}.

Similar to T$_{\rm eff}$, surface gravity (log g) values were determined for
each star using photometric temperatures and absolute bolometric magnitudes
(M$_{\rm bol.}$).  The bolometric magnitude values were calculated using the
corrections provided by \citet{Alonso99} and a distance modulus of 
(m$-$M)$_{\rm V}$ = 13.7 \citep{vandeVen06}.  Final surface gravity values were
determined using the relation,
\begin{eqnarray}
log(g_{*})=0.40(M_{bol.}-M_{bol.\sun})+log(g_{\sun})+ \\
4[log(\frac{T}{T_{\sun}})]+log(\frac{M}{M_{\sun}}) \nonumber,
\end{eqnarray}
where we assumed a mass of 0.8 M$_{\rm \odot}$ for all stars.  Although a mass
of 0.8 M$_{\rm \odot}$ is a reasonable assumption for most cluster RGB stars, 
Figure~\ref{fig:cmd_plot} shows that some of our targets may be lower mass AGB 
or red HB stars.  \citet{McDonald11} showed that $\omega$ Cen RGB stars lose 
$\sim$ 25$\%$ of their mass between the RGB and AGB sequences so a significant
fraction of our targets may have masses closer to 0.6 M$_{\rm \odot}$.  
Fortunately, the photometric surface gravity estimate shown above is only 
sensitive to the log of the mass difference, and we have restricted the present
analysis to only include \ion{Fe}{1} lines, which are not strongly pressure
sensitive in the stars targeted here.  Assuming the difference between RGB
and AGB masses does not exceed $\sim$0.2 M$_{\rm \odot}$, we adopt a log(g)
uncertainty of 0.15 (cgs).

For most stars in our analysis, the available \ion{Fe}{1} lines are too strong
to tightly constrain the microturbulence ($\xi$$_{\rm mic.}$).  Therefore, we 
determined an empirical relation between T$_{\rm eff}$ and $\xi$$_{\rm mic.}$
as,
\begin{equation}
\xi_{\rm mic.} = -6.46\times10^{\rm -4}T_{\rm eff} + 4.66,
\end{equation}
using data from \citet{Johnson10} that spanned the same T$_{\rm eff}$ range as
the stars presented here.  The adopted microturbulence calibration has a 
dispersion of 0.17 km s$^{\rm -1}$.

Since the photometric color-temperature relation is dependent on metallicity
and the surface gravity and microturbulence estimates require accurate
temperature measurements, generating a global model atmosphere solution for
each star is an iterative process.  Therefore, we initially determined 
T$_{\rm eff}$, log(g), and $\xi$$_{\rm mic.}$ for each star assuming [Fe/H] = 
$-$1.7, and iteratively solved for all four parameters as the metallicity
determination improved (see also $\S$3.3).  We interpolated within the ATLAS9
grid of model atmospheres from \citet{Castelli03}, and in general a stable 
solution was reached after $\sim$ 3 iterations.  The adopted model atmosphere
parameters, [Fe/H] values, photometry, star identifiers, and coordinates are 
provided in Table 1.

\subsection{Membership and Radial Velocities}

Although cluster membership probabilities were provided by \citet{vanLeeuwen00}
based on proper motion measurements, we verified membership using radial 
velocity measurements as well.  Radial velocities were determined using the 
first exposure of each configuration and comparing it with a synthetic 
reference spectrum at rest velocity that matched the resolution and sampling
of the data.  The observed radial velocity values were determined using the 
\emph{XCSAO} routine from \citet{Kurtz98}, and heliocentric corrections were
calculated using IRAF's \emph{rvcor} routine.  The final heliocentric radial 
velocities and \emph{XCSAO} error measurements for each star are provided in
Table 1.

We found a mean heliocentric radial velocity of $+$232.5 km s$^{\rm -1}$ for
$\omega$ Cen with a dispersion of 12.8 km s$^{\rm -1}$.  These values are in
good agreement with previous work \citep[e.g.,][]{Reijns06,Sollima09}.  A 
direct comparison of stars in common between this work and 
\citet{Reijns06} gives a mean velocity difference, in the sense of the present
work minus the literature, of $+$1.0 km s$^{\rm -1}$ with a dispersion of 
1.7 km s$^{\rm -1}$ (55 stars).  A similar comparison with \citet{Sollima09}
gives a mean difference of $+$2.2 km s$^{\rm -1}$ and a dispersion of 2.9 
km s$^{\rm -1}$ (215 stars).  The data presented in Table 1 indicate a radial 
velocity range spanning $+$186.7 to $+$271.1 km s$^{\rm -1}$; however, given 
the cluster's high radial velocity relative to the background and its large 
velocity dispersion, we designate all stars in our sample as cluster 
members\footnote{We reiterate that the targets have already been selected to
have membership probabilities $>$ 70$\%$, based on the \citet{vanLeeuwen00}
proper motion analysis.}. 

\subsection{[Fe/H] Determinations}

Given the modest S/N ratios of the spectra and the small number of available
\ion{Fe}{2} lines, all iron abundance determinations were made using synthetic
spectrum fits to \ion{Fe}{1} features.  We first selected a subset of 14
\ion{Fe}{1} lines between 5140 and 5390 \AA\ that were visible in a majority of
our spectra but had typical equivalent widths lower than about 150 m\AA.  The 
spectrum synthesis line lists were augmented to include significant lines
within 5 \AA\ of each \ion{Fe}{1} feature listed in Table 2.  Due to a paucity
of high accuracy experimental log gf values for the features given in Table 2,
empirical log gf values were determined for all lines by first adopting atomic 
data from the Vienna Atomic Line Database \citep{Ryabchikova15} compilation 
and then adjusting the log gf values until a satisfactory fit to an M2FS 
daylight solar spectrum was achieved.  We adopted the \citet{Anders89} solar 
abundances for all elements\footnote{Note that the \citet{Anders89} solar 
abundances were used for consistency with \citet{Johnson10}.}.  Although MgH 
lines may be present at wavelengths bluer than $\sim$ 5200 \AA, the roughly 
4800 K temperatures and low metallicities of our stars inhibited significant 
molecular contamination.  Furthermore, we only selected \ion{Fe}{1} lines that 
were at least $\sim$ 0.5 \AA\ away from significant molecular blends.  As a 
result, the adopted line list only included atomic features.  The adopted 
atomic parameters for all \ion{Fe}{1} lines of interest are provided in Table 2.

\begin{figure}[ht!]
\includegraphics[width=\columnwidth]{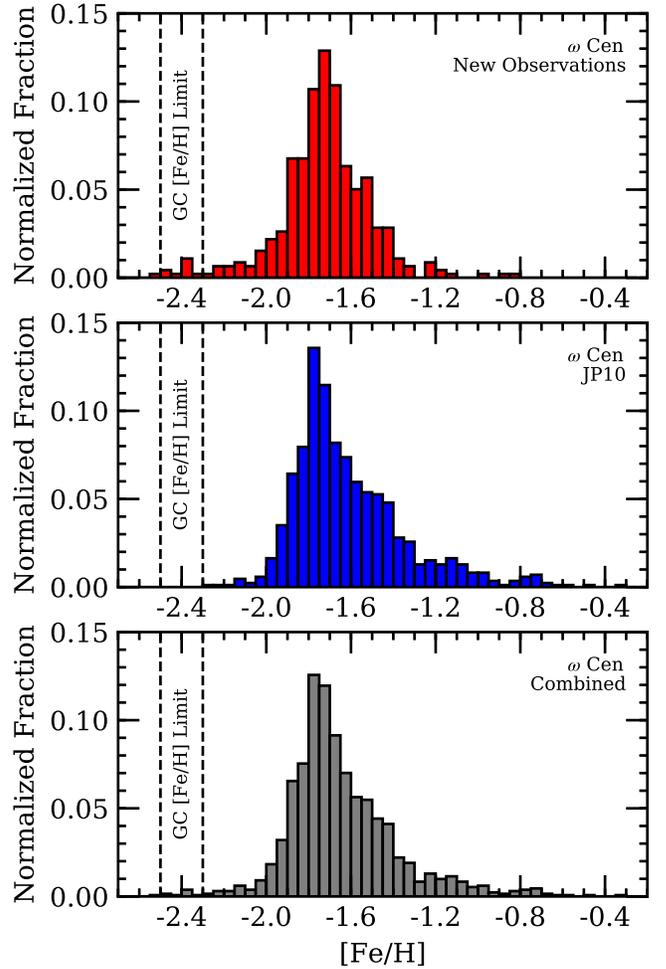}
\caption{The top, middle, and bottom panels illustrate binned metallicity
distribution functions for the new observations, the
\citet[][JP10]{Johnson10} measurements, and the combined data sets.  Note
that the biased sample presented here finds fewer stars with [Fe/H] $>$ $-$1.5
but more stars with [Fe/H] $<$ $-$2.0.  The new observations indicate an 
apparent population of very metal-poor stars that peak near [Fe/H] $\sim$ 
$-$2.4.  These stars are near or below the globular cluster metallicity floor 
found in the Milky Way and other galaxies.}
\label{fig:mdf_plot}
\end{figure}

As mentioned previously, an initial estimate of T$_{\rm eff}$, log(g), and
$\xi$$_{\rm mic.}$ was made for each star assuming [Fe/H] = $-$1.7.  For the
first iteration, the T$_{\rm eff}$, log(g), and $\xi$$_{\rm mic.}$ values were
held fixed while model atmospheres and synthetic spectra were generated for
[Fe/H] values ranging from $-$2.75 to $-$0.5, in 0.01 dex steps, using the LTE
line analysis code MOOG \citep{Sneden73}.  The fits for each line were visually
inspected to identify and remove lines affected by spectrum flaws, to set 
appropriate smoothing/broadening parameters, and to remove extreme outliers 
($\ga$ 0.3 dex from the mean).  The [Fe/H] abundance that minimized the fitting
residuals between the synthetic and observed spectra was selected as the new 
metallicity parameter, which enabled an update to the T$_{\rm eff}$, log(g), 
and $\xi$$_{\rm mic.}$ values.  This process was repeated until the difference
in mean [Fe/H] between consecutive runs decreased to $<$ 0.03 dex, and then
a final iteration was performed.  The adopted [Fe/H] abundances for each star
are provided in Table 1 and also summarized as a histogram in 
Figure~\ref{fig:mdf_plot}.

The targets were selected to have minimal overlap with other high resolution 
spectroscopic studies; however, the present sample does have 8 stars in 
common with \citet{Marino11}.  A comparison between the two studies indicates 
good agreement with a mean offset in [Fe/H] of 0.0 dex and a dispersion of 
0.13 dex, which is comparable to the mean line-to-line [Fe/H] dispersion 
reported in Table 1 for our analysis.  We also have 52 stars in common with the
low resolution (R $\sim$ 2000) spectroscopic analysis by \citet{An17} that 
shows a mean offset, in the sense of the present work minus the literature
value, of $+$0.12 dex and a dispersion of 0.15 dex.  We attribute the worse 
agreement to the order of magnitude lower resolution in the \citet{An17} data.

The [Fe/H] error values ($\Delta$[Fe/H]) provided in Table 1 take into account 
several factors added in quadrature, including the standard error of the mean 
[Fe/H] value and uncertainties related to temperature ($\Delta$T$_{\rm eff}$ = 
100 K), surface gravity ($\Delta$log(g) = 0.15 cgs), model atmosphere 
metallicity ($\Delta$[M/H] = 0.15 dex), and microturbulence 
($\Delta$$\xi$$_{\rm mic.}$ = 0.13 km s$^{\rm -1}$) errors.  However, since we 
exclusively used \ion{Fe}{1} lines the error budget is dominated by the 
T$_{\rm eff}$, $\xi$$_{\rm mic.}$, and measurement uncertainties, which also 
includes log gf and continuum normalization and/or fitting errors.  Typical 
uncertainties are approximately 0.14 dex in [Fe/H].  Since corrections for 
departures from local thermodynamic equilibrium (LTE) are not available for 
several of the lines used here, all [Fe/H] measurements assume LTE.  
Fortunately, calculations from several authors \citep{Bergemann12,Lind12,
Mashonkina16} indicate that absolute \ion{Fe}{1} non-LTE corrections should be 
$\la$ 0.05-0.10 dex in magnitude for the parameter space analyzed here.  
Relative differences between stars with similar temperatures should be even
smaller.

\section{Results and Discussion}

As mentioned in $\S$1, one of $\omega$ Cen's most intriguing properties is
its large metallicity spread.  Numerous authors have used large sample 
spectroscopic \citep{Norris95,Norris96,Suntzeff96,Villanova07,Johnson08,
Johnson10,Marino11,Pancino11,Pancino11_mr,Villanova14,An17,Mucciarelli18} 
and photometric \citep{Lee99,Hilker00,Pancino00,Rey04,Sollima05,Calamida09,
Bellini10,Calamida17} surveys to show that the cluster has at least 5-6 
populations with unique [Fe/H] values.  Peaks in the metallicity distribution 
function are generally found near [Fe/H] $\sim$ $-$1.75, $-$1.50, $-$1.20, and 
$-$0.70 (see also Figure~\ref{fig:mdf_plot}), and tend to cluster on 
well-defined RGB sequences.  \citet{Pancino11} claim to find a more metal-poor
population, rather than simply a metal-poor tail, near [Fe/H] $\sim$ $-$2, 
which suggests that additional minority populations of very metal-poor stars 
may still be awaiting discovery.

Finding cluster members with [Fe/H] $\la$ $-$2.3 would be important for linking
$\omega$ Cen's formation to a dwarf galaxy environment because such stars are 
not expected in globular clusters.  A nearly universal globular cluster 
metallicity floor exists near [Fe/H] $\sim$ $-$2.3 to $-$2.5 for Local Group
galaxies \citep[e.g.,][]{Beasley19} as well as the Milky Way 
\citep[e.g.,][]{Simpson18}, and $\omega$ Cen's retrograde, but confined to the
plane \citep[e.g.,][]{Majewski12}, orbit precludes capturing metal-poor stars
from the Galactic halo.  Therefore, very metal-poor stars found in $\omega$ Cen
today would likely have originated as a field population in the cluster's 
progenitor system.  If $\omega$ Cen formed in an environment similar to the 
dwarf galaxies observed today, then knowledge of the cluster's [Fe/H] floor 
could help place constraints on the metallicity distribution function of its 
host galaxy.

\begin{figure*}[ht!]
\includegraphics[width=\textwidth]{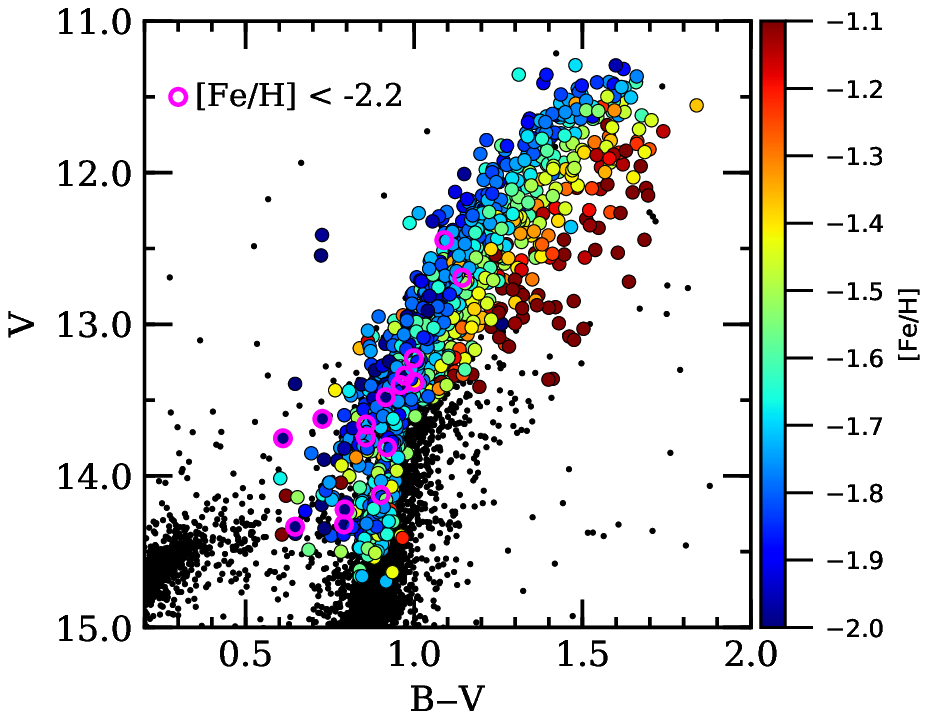}
\caption{An $\omega$ Cen V versus B$-$V color-magnitude diagram is
shown using the photometric data from \citet[][black circles]{vanLeeuwen00}.
The large filled circles are color coded by [Fe/H] using spectroscopic
abundances from \citet{Johnson10} and the present work.  The large open magenta
circles indicate stars in either study with [Fe/H] $<$ $-$2.2 dex.  Note that
the color gradient is saturated at [Fe/H] = $-$2 dex on the metal-poor end and
[Fe/H] = $-$1.1 dex on the metal-rich end.}
\label{fig:cmd_feh}
\end{figure*}

\subsection{Very Metal-poor Stars in $\omega$ Cen}

In Figure~\ref{fig:mdf_plot} we compare the biased metallicity distribution
function derived here against the relatively unbiased distribution found by
\citet{Johnson10}.  The new observations clearly confirm the presence of the
dominant metal-poor population near [Fe/H] $\sim$ $-$1.7 and the intermediate
metallicity group at [Fe/H] $\sim$ $-$1.5, and as expected from our selection
procedure we find far fewer stars with [Fe/H] $\ga$ $-$1.3.  
Figure~\ref{fig:cmd_feh} further shows that our [Fe/H] determinations follow
the expected correlation between metallicity and RGB color, and that most of
the stars in our sample with [Fe/H] $\ga$ $-$1.3 lie on the AGB or red HB
sequences.  

We find a possible metal-poor population near [Fe/H] $\sim$ $-$2.1 that appears
separate from the dominant group at [Fe/H] $\sim$ $-$1.7 (see also $\S$ 4.1.1),
and we suspect that these RGB stars are part of the same metal-poor faction 
identified by \citet{Pancino11} on the SGB.  Interestingly, we also find a new 
population of 11 stars with metallicities below the [Fe/H] = $-$2.26 limit set 
by \citet{Johnson10}.  These stars represent 2.8$\%$ of our sample and span from
[Fe/H] = $-$2.30 to $-$2.52.  Figure~\ref{fig:cmd_feh} further shows that 
these very metal-poor stars generally fall on the blue edge of the cluster RGB,
as expected, but 2-3 of the targets could be AGB or HB stars.  The very
metal-poor stars identified here are close to the empirically determined 
globular cluster metallicity floor noted in \citet{Beasley19}.

The existence of a metal-poor tail in $\omega$ Cen is already a unique trait 
among iron-complex clusters.  The metallicity distribution functions for all 
iron-complex clusters except $\omega$ Cen possess a sharp rise at low 
metallicity, one or more populations at higher metallicity, and occasionally a 
metal-rich tail.  The only other compelling case for a cluster hosting a 
metal-poor tail or a minor population at low metallicity is Terzan 5, which has
a dominant population near [Fe/H] $\sim$ $-$0.3, a secondary population at 
[Fe/H] $\sim$ $+$0.25, and a minority ($\la$ 6$\%$) metal-poor group at 
[Fe/H] $\sim$ $-$0.8 \citep{Origlia13,Massari14}.  However, the various 
Terzan 5 sub-populations do not possess the complex light element variations 
observed in $\omega$ Cen.

\begin{figure*}[ht!]
\includegraphics[width=\textwidth]{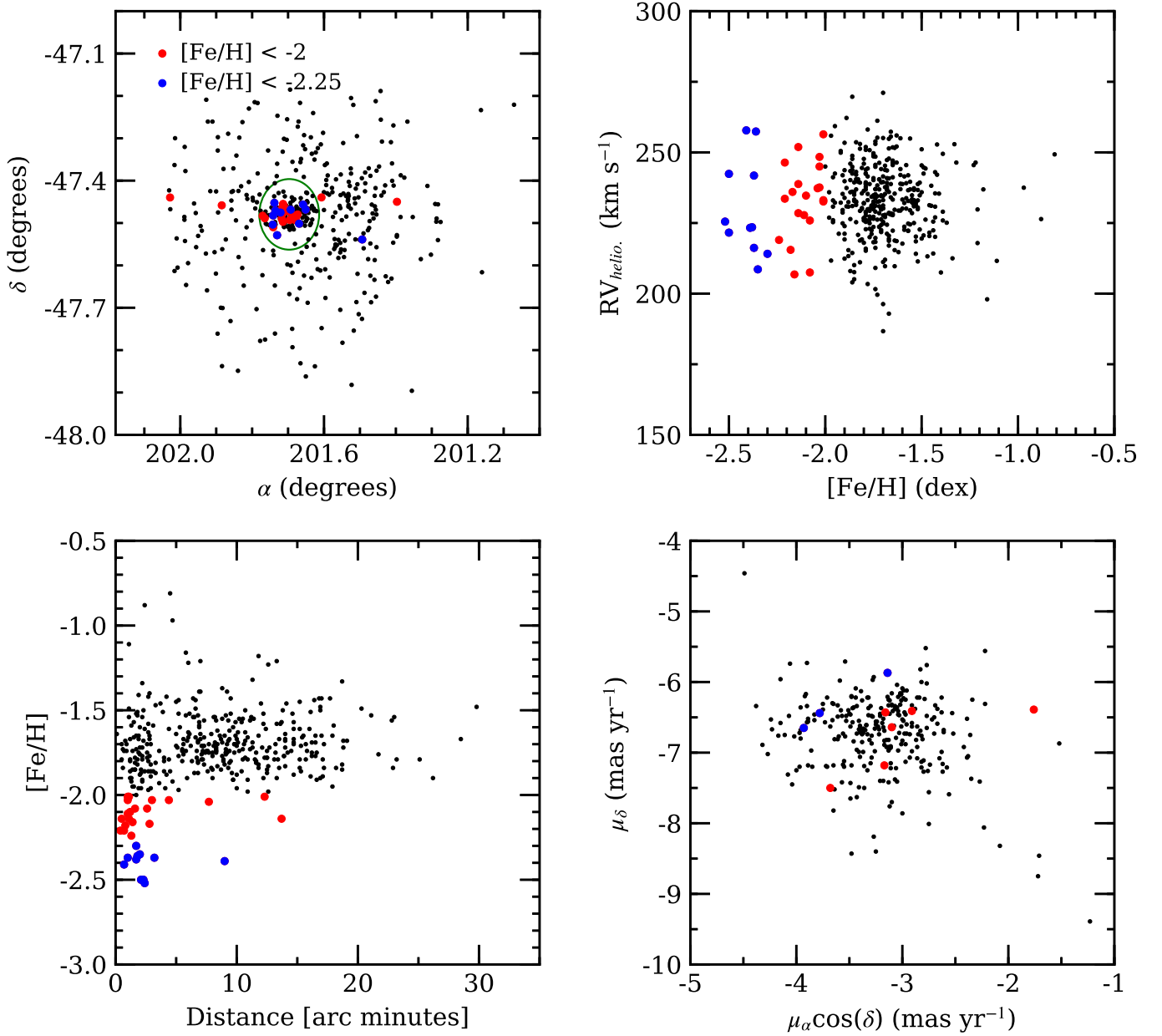}
\caption{\emph{Top left:} the black circles indicate the J2000 sky coordinates 
of all $\omega$ Cen giants observed for the present study while the large red
circles indicate stars with [Fe/H] $<$ $-$2 dex.  The most metal-poor stars
with [Fe/H] $<$ $-$2.25 are indicated as large blue circles.  The green
ellipse illustrates the cluster's approximate half-light radius (5$\arcmin$).
Although nearly all of the metal-poor stars identified in the present work
reside within the cluster half-light radius, the stars farther out are only
found within a narrow declination range.  \emph{Top right:} heliocentric radial
velocities are plotted as a function of [Fe/H] using the same symbols as in
the left panel.  The metal-poor stars do not exhibit any obvious radial
velocity peculiarities.  \emph{Bottom left:} [Fe/H] is plotted as a function 
of radial distance from the cluster center.  Note that the stars with [Fe/H]
$<$ $-$2 appear to be centrally concentrated.  \emph{Bottom right:} Gaia DR2
proper motion values are plotted using the same symbols as the left panel.  We 
do not find the metal-poor stars to exhibit any systematic proper motion 
differences relative to the higher metallicity stars.}
\label{fig:mpoor_plot}
\end{figure*}

Other than metallicity, do the most metal-poor stars in $\omega$ Cen possess 
any unusual properties?  Figure~\ref{fig:mpoor_plot} compares the radial and
spatial distributions, radial velocities, and Gaia DR2 proper motions for the 
stars analyzed in the present study.  Visual inspection of the two left panels 
in Figure~\ref{fig:mpoor_plot} indicate that: (1) stars with [Fe/H] $<$ $-$2 
span a broad range in right ascension but only populate a narrow range in 
declination and (2) the most metal-poor stars may be very centrally 
concentrated.  First addressing the spatial distribution, we note that the 
dispersion in declination for stars with [Fe/H] $>$ $-$2 is 0.128 degrees
while the dispersions for stars with $-$2.25 $\leq$ [Fe/H] $\leq$ $-$2 and 
[Fe/H] $<$ $-$2.25 are 0.023 and 0.029 degrees, respectively.  Similarly,
two-sided Kolmogorov-Smirnov (KS) tests comparing the declination distributions
of all stars versus first those with [Fe/H] $<$ $-$2 and then those with [Fe/H]
$<$ $-$2.25 returned $p$-values\footnote{We adopt the common convention that
a $p$-value $<$ 0.05 is sufficient to reject the null hypothesis.} of 0.002 and
0.095, respectively.  From these data we can conclude that there is marginal
evidence indicating that the most metal-poor $\omega$ Cen stars may 
be confined to a more narrow plane than the majority of cluster stars.
However, a larger sample size is required to draw any strong conclusions.

Regarding the radial distributions, Figure~\ref{fig:mpoor_plot} shows that 
nearly all of the stars with [Fe/H] $<$ $-$2 are projected within 12$\arcmin$ 
of the cluster core and 84$\%$ are within 3$\arcmin$.  Similarly, all of the 
stars with [Fe/H] $<$ $-$2.25 are within 10$\arcmin$ and 81$\%$ are inside of 
3$\arcmin$.  These results match previous observations by \citet{Johnson10}
who noted that 88$\%$ of the stars in their sample with [Fe/H] $<$ $-$2 were
found within 5$\arcmin$ of the cluster core.  The combination of the present
work with the larger sample of \citet{Johnson10} hints that the
most metal-poor stars in $\omega$ Cen may be mostly confined to within
a half-light radius (5$\arcmin$).  A two-sided KS test comparing the radial
distributions of stars with [Fe/H] $>$ $-$2 against those with [Fe/H] $<$ $-$2
returned a $p$-value of 2.91$\times$10$^{-10}$ while a similar test comparing
the radial distributions of stars with [Fe/H] $>$ $-$2 and [Fe/H] $<$ $-$2.25
returned a $p$-value of 8.66$\times$10$^{-5}$.  The extremely low $p$-values
indicate a low probability that stars with [Fe/H $<$ $-$2 follow the same 
radial distribution as their higher metallicity counterparts.  We did 
not detect any further differences between the most metal-poor stars and any 
other populations when examining radial velocity or proper motion 
distributions; however, larger sample sizes are required to draw any
firm conclusions.

\subsubsection{Do the Very Metal-poor Stars Form Distinct Populations?}

An important question when examining the metallicity distribution function of
$\omega$ Cen is whether the stars with [Fe/H] $<$ $-$2 form distinct 
populations or simply an extended tail.  As noted previously, the binned 
distributions shown in Figure~\ref{fig:mdf_plot} indicate possible ``peaks" 
near [Fe/H] $\sim$ $-$2.1 and $-$2.4, but the sample sizes are small.  To help 
determine whether the most metal-poor stars are separate populations, we 
utilize the Gaussian mixture model (GMM) method described in \citet{Ashman94} 
via the implementation outlined in \citet{Muratov10} to examine: (1) if the 
peak near [Fe/H] = $-$2.1 is distinct from the dominant population near 
[Fe/H] = $-$1.7; and (2) if the groups at [Fe/H] = $-$2.1 and $-$2.4 are 
separate or part of a broad unimodal population.

The modality analysis described in \citet{Muratov10} includes an algorithm for 
a likelihood ratio test that determines whether the sum of two Gaussian 
profiles provides an improvement over fitting a unimodal Gaussian
distribution.  Note that we have adopted the homoschedastic case (equal 
variance) for all GMM tests, which is a reasonable assumption if the [Fe/H] 
spread within each population is dominated by equivalent measurement errors.
Since the likelihood ratio test assumes Gaussian distributions, 
\citet{Ashman94} and \citet{Muratov10} also include a dimensionless mode 
separation statistic defined as,
\begin{equation}
D\equiv \frac{\left |\mu_{1}-\mu_{2}  \right |}{\sqrt{(\sigma_{1}-\sigma_{2})/2}},
\end{equation}
where D $>$ 2 is required for a clear separation between two modes when
the mean ($\mu$) and dispersion ($\sigma$) values for two populations are known.
\citet{Muratov10} also note that a negative kurtosis is a ``necessary but
not sufficient condition of bimodality".  However, since the number of very 
metal-poor stars in our sample is small, especially relative to the dominant 
population at [Fe/H] $\sim$ $-$1.7, we do not include kurtosis measurements 
here.

For the first case listed above, we compiled a list of stars with $-$2.25 
$\leq$ [Fe/H] $\leq$ $-$1.60 from Table 1 and ran the \citet{Muratov10} 
algorithm to determine whether a bimodal distribution was an improvement over 
fitting a single Gaussian profile.  The \citet{Muratov10} test found that the
data were better fit assuming two Gaussian distributions with means of 
[Fe/H] = $-$2.07 and $-$1.76 ($\sigma$ = 0.09 dex) over a single Gaussian with 
a mean of [Fe/H] = $-$1.78 and a larger dispersion ($\sigma$ = 0.11 dex) at 
more than the 99.9$\%$ level.  A bootstrap analysis yielded a separation 
statistic of D = 3.38 $\pm$ 0.41, which provides a high level of confidence 
that the peak at [Fe/H] = $-$2.07 represents a legitimate population rather 
than a metal-poor tail for the [Fe/H] = $-$1.76 group.

For the second case, we examined the [Fe/H] distribution for all stars in 
Table 1 with [Fe/H] $<$ $-$2 to determine if these objects form two narrow 
populations or one broad distribution.  We determined that a unimodal 
distribution was ruled out at the more than 99.6$\%$ level, and that two 
populations were preferred having means of [Fe/H] = $-$2.4 and $-$2.1 with
dispersions of 0.07 dex.  The bootstrap mode separation statistic was 
calculated to be D = 4.37 $\pm$ 0.8, which suggests that if two populations
are present below [Fe/H] = $-$2 then their mean metallicities are 
well-separated. 

\subsection{Origin of the Very Metal-poor Stars}

If the stars in $\omega$ Cen with [Fe/H] $<$ $-$2, and especially those with
[Fe/H] $<$ $-$2.25, are truly distinct from the other cluster populations,
then their presence may reveal important details about the cluster's early 
formation history.  The available evidence suggests several possible origins
for $\omega$ Cen's very metal-poor stars, including: (1) early star formation 
in the unenriched primordial cloud, (2) the capture of surrounding field stars, 
assuming the cluster formed in a dwarf galaxy, or (3) an early merger between 
$\omega$ Cen and a pre-existing but metal-poor sub-structure.  

\subsubsection{Early Star Formation}

The primordial cloud scenario is the most straight-forward, and would imply 
that the very metal-poor stars trace the original composition of the molecular
cloud from which $\omega$ Cen formed.  Data from \citet{Johnson10} and 
\citet{Pancino11} indicate that stars with [Fe/H] $<$ $-$2 may have 
reduced light element abundance variations compared to their more
metal-rich counterparts, which might be consistent with an early
proto-cluster environment that had not yet been polluted by sources such as 
intermediate mass AGB stars.  However, if $\omega$ Cen actually has two
populations of very metal-poor stars at [Fe/H] $\sim$ $-$2.1 and $-$2.4, as
is suggested in Figure~\ref{fig:mdf_plot} and $\S$ 4.1.1, then the cluster 
could have experienced at least two early small bursts of star formation.
Detailed light and heavy element abundance measurements for the [Fe/H] $\sim$ 
$-$2.4 stars identified here would help determine if the two groups are 
chemically similar.

Examining the published metallicity distribution functions
of other iron-complex clusters, we note that aside from Terzan 5 only M 54
appears as a possible candidate to host a metal-poor tail.  However, the most 
metal-poor stars in this system were clearly polluted, as evidenced by their 
large (anti-)correlated light element abundance variations 
\citep[e.g.,][]{Carretta10_sag}, and would therefore not represent a pristine 
population.  Similarly, we note that monometallic clusters have significant 
populations of ``primordial" stars that appear to be almost entirely polluted 
by supernovae, but these stars do not have different [Fe/H] abundances from 
other cluster members.  Therefore, either $\omega$ Cen and Terzan 5 were unique
in their ability to retain stars that formed before the first major star 
formation and light element enrichment episode, or their very metal-poor 
populations have a different origin.

\subsubsection{Captured Field Stars}

The captured field star scenario is intriguing because it could provide a 
direct link between present day iron-complex clusters and dwarf galaxies.
However, if $\omega$ Cen formed within a dwarf galaxy should we 
expect such a system to host many very metal-poor stars?  We can draw two 
reasonable conclusions based on $\omega$ Cen's properties alone.  First, 
$\omega$ Cen's mass of $\sim$ 4$\times$10$^{\rm 6}$ M$_{\rm \sun}$ 
\citep{DSouza13} means that its host galaxy must have had a mass of at least
10$^{\rm 7}$-10$^{\rm 8}$ M$_{\rm \sun}$ \citep[see also][]{Bekki19,Myeong19}.
Second, the dwarf galaxy mass-metallicity relation from \citet{Kirby13} implies
that the cluster's host galaxy likely had a mean metallicity of at least 
[Fe/H] = $-$1.5 to $-$1.  A comparatively high metallicity field star 
population is qualitatively in agreement with the M 54/Sagittarius system
where the mean metallicity of M 54 ([Fe/H] $\sim$ $-$1.6) is several times
lower than the Sagittarius field \citep[e.g.,][]{Carretta10_sag}.  
Nevertheless, dwarf galaxies tend to have long metal-poor tails 
\citep[e.g.,][]{Leaman13}, and even cases such as Sagittarius that have a 
majority of stars with [Fe/H] $>$ $-$1 contain stars with metallicities as
low as $-$2.2 \citep{Chiti19}.  It therefore remains plausible that $\omega$ 
Cen could have existed in an environment that also hosted very metal-poor 
field stars.

The ability of $\omega$ Cen to capture very metal-poor stars might depend not 
only on the metallicity distribution of its host galaxy, but also the cluster's 
typical galactocentric radius and the galaxy's star formation history before 
tidal disruption.  Existing Local Group dwarf galaxies generally exhibit
negative metallicity gradients such that the mean metallicity decreases at 
larger galactocentric distances \citep[e.g.,][]{Harbeck01,Battaglia06,
Faria07,Carrera08,Kirby11,Kacharov17}.  If $\omega$ Cen resided in the core
of its host galaxy and a significant fraction of the cluster's metal-poor stars
were captured, then $\omega$ Cen's host galaxy might have had an unusually 
metal-poor core.  Alternatively, the cluster's host galaxy could have had an 
inverted metallicity gradient \citep[e.g.,][]{Wang19}.

Although $\omega$ Cen is the most massive cluster associated with the Sequoia 
\citep{Myeong19} or Gaia-Enceladus \citep{Massari19} merger events, that does 
not necessarily mean the cluster always (or ever) resided in its host galaxy's 
center.  For example, M 54 is the most massive cluster associated with 
Sagittarius and presently lies near the galaxy's core, but the cluster could 
have formed outside the galaxy nucleus and drifted to the center via dynamical 
friction \citep[e.g.,][]{Bellazzini08}.  Furthermore, the Fornax dwarf galaxy 
has five globular clusters \citep{Shapley38,Hodge61} and all except one (Fornax
4), including the most massive and metal-poor cluster Fornax 3 
\citep{deBoer16}, reside well outside the core radius.  One might expect that 
if $\omega$ Cen were in a configuration similar to that of Fornax 3 it would 
have a reasonable chance of capturing metal-poor field stars.

Regardless of where $\omega$ Cen may have resided within a host galaxy, it 
could not capture metal-poor field stars unless they were present in 
significant numbers.  Combining the present sample with \citet{Johnson10}, 
3.9$\%$ of stars in $\omega$ Cen have [Fe/H] $<$ $-$2 and about 1$\%$ have
[Fe/H] $<$ $-$2.3.  For comparison, \citet{Starkenburg10} showed that the 
fractions of stars with [Fe/H] $<$ $-$2.5 were 1$\%$, 8$\%$, 8$\%$, and 33$\%$
for the Fornax, Carina, Sculptor, and Sextans Local Group dwarf galaxies,
respectively.

These observations suggest that $\omega$ Cen's host galaxy probably had a 
higher fraction of metal-poor stars than Fornax, especially if all $\omega$
Cen stars with [Fe/H] $<$ $-$2 were captured.  In contrast, the metallicity
distribution of Sextans may be too metal-poor since we did not find any stars
with [Fe/H] $\la$ $-$2.5.  However, we investigated the color-color 
distribution of stars in $\omega$ Cen using recent Sky Mapper \citep{Wolf18}
photometry and found a small number of stars that may have [Fe/H] $<$ $-$2.5 
to $-$3.  A galaxy like Sculptor, but with the globular
cluster specific frequency of Fornax, may serve as a reasonable model of 
$\omega$ Cen's host galaxy since it has: (1) a mean [Fe/H] value that is higher
than $\omega$ Cen, (2) a small but significant population of field stars down 
to [Fe/H] $\sim$ $-$3.0, and (3) few stars with [Fe/H] $\ga$ $-$1 
\citep[e.g.,][]{Kirby09,Starkenburg10}.  The small ($\sim$ 10 km s$^{\rm -1}$) 
velocity dispersion of Fornax globular clusters and field stars 
\citep[e.g.,][]{Hendricks14} would also provide a model that is conducive to 
tidal capture, but it remains to be seen whether such a scenario could be 
reconciled with the observed central concentration of very metal-poor stars in 
$\omega$ Cen.

\subsubsection{Merging Sub-clumps}

Globular clusters are likely the end results of complicated hierarchical
merging processes \citep[e.g.,][]{Bonnell03,Smilgys17}, and in this sense
the rarity of very metal-poor stars among iron-complex clusters could be a
result of this stochastic process.  Sub-clump and cluster-cluster mergers have 
been invoked to explain several globular cluster properties in the past
\citep[e.g.,][]{vandenBergh96,Lee99,Carretta10_1851,Bekki19}, and we posit that
such a model may explain the existence of the metal-poor populations in 
$\omega$ Cen, and possibly Terzan 5, as well.

In this model, the early $\omega$ Cen environment would have been subject to
intense star formation and molecular cloud and/or cluster mergers.  As a 
result, the main $\omega$ Cen structure could have coalesced with a 
sub-structure that was either in the process of forming or already fully 
formed, but that had a very low metallicity and not enough time to experience 
significant light element self-enrichment.  Since the most metal-poor stars in 
$\omega$ Cen do not reach below the empirical globular cluster limit of [Fe/H] 
$\sim$ $-$2.5 and Figure~\ref{fig:mdf_plot} hints at two different populations 
with [Fe/H] $<$ $-$2 rather than a continuous metal-poor tail, a merging 
sub-structure scenario may be favored over a field star capture model.  If the 
merging sub-structures were also relatively massive, they might naturally fall 
to the cluster center and help explain the central concentration of stars with 
[Fe/H] $<$ $-$2 shown in Figure~\ref{fig:mpoor_plot}.

Although the stars with [Fe/H] $<$ $-$2 constitute only a small fraction of 
$\omega$ Cen's mass, we did not detect any unusual kinematic properties for
these populations.  As a result, the sub-clump merger origin for very 
metal-poor stars may only be plausible if their initial kinematic signature 
was erased by dynamical evolution or was very similar to the existing 
$\omega$ Cen proto-cluster.  

\section{Summary}

We present [Fe/H] measurements, based on high resolution M2FS spectra, for 395 
giants in the massive globular cluster $\omega$ Cen.  The targets were chosen
to reside on the blue half of the RGB in an effort to find the most metal-poor
stars in the cluster.  Previous attempts identified a metal-poor population 
with [Fe/H] $\sim$ $-$2.1 but failed to find any stars with [Fe/H] $\la$ 
$-$2.25.  However, we have identified 11 new stars with metallicities ranging
from [Fe/H] = $-$2.30 to $-$2.52, which places these stars near the 
empirical globular cluster metallicity floor observed among the Milky Way and
other Local Group galaxies.  

The metal-poor stars identified here and in previous studies appear to be very
centrally concentrated, and may also be confined to a narrow declination plane.
However, these stars do not appear to exhibit any peculiar kinematic properties
that distinguish them from the more metal-rich populations in $\omega$ Cen.
We examine three possible scenarios in which $\omega$ Cen could form or capture
stars that are significantly more metal-poor than its dominant population at
[Fe/H] $\sim$ $-$1.7.  Specifically, the very metal-poor stars could: (1) trace
very early star formation in the cluster, (2) have originated as captured field
stars from $\omega$ Cen's original host galaxy, or (3) represent an early 
merger event between $\omega$ Cen and metal-poor sub-clumps.  Although none of 
these scenarios is clearly favored, the merging sub-clump model may have the 
greatest chance for describing the paucity of very metal-poor stars in other 
clusters, the possibly reduced light element spread for stars with [Fe/H] $<$ 
$-$2, and the observed central concentration of these populations.

\acknowledgments

CIJ and AKD gratefully acknowledge support from the Scholarly Studies Program
of the Smithsonian Institution.  EWO, MGW, and MM acknowledge support from the 
National Science Foundation under grants AST-1815767, AST-1813881, and 
AST-1815403.

%

\vspace{5mm}
\facilities{Magellan(M2FS)}






\bibliography{references}{}
\bibliographystyle{aasjournal}



\begin{longrotatetable}
\begin{deluxetable*}{ccccccccccccccc}
\tablecaption{Sky Coordinates, Photometry, Model Atmosphere Parameters, Iron
Abundances, Errors, and Radial Velocities}
\tablewidth{700pt}
\tabletypesize{\scriptsize}
\tablehead{
\colhead{LEID}	&
\colhead{RA}	&
\colhead{DEC}	&
\colhead{V}	&
\colhead{J}	&
\colhead{K$_{\rm S}$}	&
\colhead{T$_{\rm eff}$}	&
\colhead{log(g)}	&
\colhead{[Fe/H]}	&
\colhead{$\xi$$_{\rm mic.}$}	&
\colhead{$\sigma$/$\sqrt{N}$}	&
\colhead{N}	&
\colhead{$\Delta$[Fe/H]}	&
\colhead{RV$_{\rm Helio.}$}	&
\colhead{Error}	\\
\colhead{}	&
\colhead{(degrees)}      &
\colhead{(degrees)}      &
\colhead{(mag.)}      &
\colhead{(mag.)}      &
\colhead{(mag.)}      &
\colhead{(K)}      &
\colhead{(cgs)}      &
\colhead{(dex)}      &
\colhead{(km s$^{\rm -1}$)}      &
\colhead{(dex)}      &
\colhead{}	&
\colhead{(dex)}      &
\colhead{(km s$^{\rm -1}$)}      &
\colhead{(km s$^{\rm -1}$)}	
}
\startdata
4013	&	201.44247	&	$-$47.18851	&	13.982	&	12.111	&	11.511	&	4960	&	2.03	&	$-$1.49	&	1.46	&	0.05	&	14	&	0.12	&	220.6	&	0.4	\\
4022	&	201.69444	&	$-$47.18516	&	14.635	&	12.745	&	12.158	&	4953	&	2.28	&	$-$1.43	&	1.46	&	0.03	&	14	&	0.12	&	220.1	&	0.6	\\
6012	&	201.52474	&	$-$47.20504	&	13.685	&	11.881	&	11.278	&	5035	&	1.94	&	$-$1.95	&	1.41	&	0.05	&	12	&	0.14	&	232.2	&	0.3	\\
7010	&	201.45608	&	$-$47.21242	&	14.662	&	12.858	&	12.307	&	5097	&	2.36	&	$-$1.72	&	1.37	&	0.03	&	12	&	0.14	&	236.9	&	0.7	\\
7022	&	201.92768	&	$-$47.20907	&	14.338	&	12.399	&	11.790	&	4874	&	2.12	&	$-$1.85	&	1.51	&	0.06	&	12	&	0.14	&	205.1	&	0.4	\\
8004	&	201.07113	&	$-$47.22082	&	14.508	&	12.687	&	12.007	&	4928	&	2.22	&	$-$1.48	&	1.48	&	0.02	&	13	&	0.12	&	244.3	&	0.3	\\
8022	&	201.66668	&	$-$47.21645	&	14.034	&	12.085	&	11.454	&	4840	&	1.99	&	$-$1.74	&	1.53	&	0.05	&	13	&	0.14	&	244.9	&	0.3	\\
8026	&	201.78482	&	$-$47.21673	&	14.315	&	12.397	&	11.785	&	4894	&	2.12	&	$-$1.77	&	1.50	&	0.04	&	13	&	0.14	&	229.3	&	0.5	\\
\enddata
\end{deluxetable*}
\end{longrotatetable}

\begin{deluxetable*}{ccc}
\tablecaption{\ion{Fe}{1} Line List}
\tablewidth{0pt}
\tablehead{
\colhead{Wavelength}	&
\colhead{E.P.}	&
\colhead{log gf}	\\
\colhead{(\AA)}	&
\colhead{(eV)}	&
\colhead{}
}
\startdata
5141.739	&	2.422	&	$-$2.00	\\
5145.094	&	2.196	&	$-$3.06	\\
5150.839	&	0.989	&	$-$3.07	\\
5159.058	&	4.280	&	$-$0.65	\\
5198.711	&	2.221	&	$-$1.97	\\
5213.344	&	4.383	&	$-$2.18	\\
5216.274	&	1.607	&	$-$2.15	\\
5242.491	&	3.632	&	$-$0.84	\\
5243.776	&	4.253	&	$-$0.90	\\
5247.050	&	0.087	&	$-$4.65	\\
5288.527	&	3.692	&	$-$1.54	\\
5321.108	&	4.431	&	$-$1.24	\\
5322.041	&	2.277	&	$-$2.85	\\
5379.574	&	3.692	&	$-$1.42	\\
\enddata
\end{deluxetable*}

\end{document}